\documentclass[english,12pt,twoside]{article}
\usepackage{amssymb}
\usepackage{amsmath}
\usepackage[dvips]{graphicx}
\begin{document}

\date{}
\title{Transitivity vs. Intransitivity \\in decision making process.\\\small (An example in quantum game theory)}
\markboth{Marcin Makowski}{Transitivity vs. Intransitivity}
\author{Marcin Makowski\footnote{makowski.m@gmail.com}\\[1ex]\small
Institute of Mathematics, University of Bia\l ystok,\\\small
Akademicka 2, PL-15424, Bia{\l}ystok,
Poland}
\maketitle\begin{abstract}
 We compare two different ways of quantization a simple sequential game Cat's Dilemma in the context of the debate on intransitive and transitive preferences. This kind of analysis can have essential meaning for the research on the artificial intelligence (some possibilities are discussed). Nature has both properties transitive and intransitive and maybe quantum models can be more able to capture this dualism than classical one. We also present electoral interpretation of the game.
\end{abstract}
\section{Introduction}
In the work we concentrate on a quantitative analysis of two quantum modifications (quantization and prequantization) of a very simple game against Nature which was presented and analyzed in \cite{r1}. In this context we analyze an important aspect of games -- intransitive strategies. It is a theme that deserves a thorough analysis in the language of quantum game theory \cite{r2, r3, r4, r5, r6, r7, r8, r9, r10}. A more profound analysis of intransitive orders can have importance everywhere where the problem of choice behavior is considered (for those investigating mind performance or the construction of thinking machines -- artificial intelligence).
The geometric interpretation presented in this paper enables us to track the evolution of a hypothetical \emph{intelligence} in the process of quantization of a game. A similar analysis of the various aspects of quantum games (depending on how we model the game) may contribute to the formulation of a number of interesting proposals and observations, and have importance for research on the general properties of games in the context of quantum information theory \cite{r3Niels}.

This paper is organized as follows: Section 2 introduces the reader to the intransitivity concept. In Section 3, for completeness of this paper, we describe the classical model of Cat's Dilemma. Section 4 introduces the concept of prequantization and quantization in the Cat's Dilemma game. In Section 5 we compare the effects offered by various models of the game. Section 6 presents an electoral interpretation of the game. In Section 7 some possible application are indicated.

\section{Intransitivity}
Any relation $\succ$ existing between the elements of a certain set is called \emph{transitive} if $A\succ C$ results from the fact that $A\succ B$ and $B\succ C$ for any three elements  $A$, $B$, $C$. If this condition is not fulfilled then the relation will be called \emph{intransitive} (not
transitive).

There is a view (supported mainly by economists) that people, based on rational premises, should choose between things they like in a specific, linear order (logical order)\cite{r29}. This axiom is the fundamental element of classical theories of individual and collective choices. We come upon this understanding of rationality in everyday life, where all rankings and ratings are transitive. Intransitive relations are often perceived as something paradoxical. This is probably the consequence of transitive inference. This type of reasoning is a developmental milestone for human children, showing up as early as in ages 4 and 5 \cite{r33}. It allows children to reason that if \emph{A} is bigger than \emph{B}, and  \emph{B} is bigger than  \emph{C}, then  \emph{A} is also bigger than \emph{C}. This reasoning has been reported in primates, rats, birds and recently also in fish \cite{r33}. This feature is defined by biologists as the essence of logical thinking. It must be noted here that there is no evidence that animals use it consciously. However, this does not change the fact that from an evolutionary point of view, transitive inference saves a lot of time and energy and seems to us to be quite natural in daily decision-making. 

One of the main arguments put forward by many experts, proving the irrationality of preferences which violate transitivity, is the so-called ``money pump'' \cite{r23}. Suppose that an agent has a clearly defined intransitive preference: he/she prefers $X$ to $Y$, $Y$ to $Z$, and $Z$ to $X$ and he/she is the owner of $X$. The agent is offered the opportunity to switch from $X$ to $Z$ for one dollar, and he/she accepts. He/she is then offered the opportunity to switch from $Z$ to $Y$ for one dollar, and he/she accepts. And, he/she is offered the opportunity to switch from $Y$ to $X$ for one dollar, which he/she accepts. Such a sequence of transactions would lead to the paradoxical situation in which the agent paid for the thing which was in his/her possession.

Despite the fact that intransitivity appears to be contrary to our intuition, life provides many examples of intransitive orders. They often even seem to be necessary and play a positive role, although we probably do not pay any more attention to this. Such a situation takes place, for example, in sport. Team $A$ which defeated $B$, which in turn won with $C$, can be overcome by team $C$. This makes the result unpredictable and more exciting. A similar kind of benefit is in the case of a children's game called {\itshape Stone, Scissors, Paper}. The relation used to determine which throws defeat which is intransitive -- Rock defeats Scissors, Scissors defeat Paper, but Rock loses to Paper. So the game is more interesting because there is no sign of a dominant other (you can play it not knowing in advance who will win). At this point it is worth mentioning \emph{the games in public goods} that are used as models to study the social ties and interactions in a group of individuals. It turns out that the participants of such games coexist through rock-paper-scissors dynamics \cite{r18, r19}.

The best known and socially significant example of intransitivity is Condorcet's voting paradox. Consideration of this paradox led Arrow (in the 20th century) to prove the theorem that there is no procedure of successful choice that would meet a democratic assumption \cite{r24}. Interesting examples of intransitivity are provided by probability models (Efron's dice  \cite{r34}, Penney's game \cite{r35}).

Intransitive models can explain many processes in the world of nature. Rivalry between species may be intransitive. For example, in the case of fungi, Phallus impudicus replaced Megacollybia platyphylla, M. platyphylla replaced Psathyrella hydrophilum, but P. hydrophilum replaced P. impudicus \cite{r15}. Similarly, the experiment can be explained on bees, which make intransitive choices between flowers \cite{r16}.
In chemistry, one of the best known examples of intransitive order is the so-called Belousov-Zhabotinsky reaction, in which different colors of liquid sequentially replace one another again and again. Belousov had enormous problems with the publication of his discoveries for many years, with many chemists thinking that such a reaction is impossible \cite{r30}.
Another example in the field of chemistry is the Krebs Cycle, which is the common and fundamental pathway for all biochemical energy processes. 

Many researchers believe that the essence of the transitivity preference is a rational choice, and intransitivity preferences can be formed only by negligent decisions. There are also those who are trying to reconcile intransitive preferences with rationality, recognizing that this does not always have to mean an action which is contrary to our interests \cite{r22}. We have to remember that what appears in one situation to be manifestly reasonable, in another could seriously do harm. It is very important to precisely define how we understand rationality in a specific situation, which is not always an obvious and easy task, because anyone can read it in different ways according to their own criteria. 

It turns out that, thanks to the intransitivity orders, we can get optimal results \cite{r1, r2op}.
A total rejection of intransitivity preferences as undesirable would be an unnecessary restriction of research in the field of game theory and decisions. They are still a mysterious and surprising feature of nature, of human, and not only human, thought processes, and understanding the mechanisms of their formation may be important in the search for the logic of thinking (a universal which is common to all species), as well as research on artificial intelligence. 
We might look at intransitivity orders by using a new language of modeling the decision-making algorithms, which is quantum games theory. Perhaps the quantum model describes features related to decision-making more precisely than the classical models. Quantum mechanics provide a natural explanation for the decision-making process \cite{r36}.

In the paper we take a look at two different ways of modeling the same game in quantum game theory language and, in particular, how these different approaches influence the occurrence of intransitivity preferences. 

To illustrate the problem, we will use the story of Pitts's experiments with cats mentioned in the Steinhaus diary \cite{r25}. Pitts noticed that a cat, facing the choice between fish, meat and milk prefers fish to meat, meat to milk, and milk to fish! Pitts's cat, thanks to the above-mentioned food preferences, provided itself with a balanced diet (one of the key factors needed to maintain good health).

\section{Cat's Dilemma. The classical model}
Let us assume that a cat is offered three types of food (no. 0, no. 1 and no. 2),
every time in pairs of two types, whereas the food portions are equally attractive
regarding the calories, and each one has some unique components that are necessary
for the cat's good health. The cat knows (it is accustomed to) the frequency of
occurrence of every pair of food and his strategy depends on only this frequency.
Let us also assume that the cat cannot consume both offered types of food at the
same moment, and that it will never refrain from making the choice.

 The eight ($2^3$) possible deterministic choice functions $f_k$:
 \begin{displaymath}
f_k:\{(1,0), (2,0),(2,1)\}\to \{0,1,2\},\qquad k=0,\ldots,7,
\end{displaymath} 
 are defined in Tabel \ref{Tabela1} on page \pageref{Tabela1}.
The parameters $p_k$, $k = 0,\ldots, 7$ give the frequencies of appearance of the choice function
in the nondeterministic algorithm (strategy) of the cat.

There is the following relation between the frequencies $\omega_k$, $k=0,1,2$,
 of appearance of the particular foods in a diet
and the conditional probabilities which we are interested in
(see \cite{r1}):
\begin{equation} 
\omega_k
:=P(C_k)=\sum_{j=0}^{2}P(C_{k}|B_{j})P(B_{j}),\,\, k=0,1,2,
\end{equation}
where $P(C_{k} | B_{j})$ indicates the probability of choosing the food
of number $k$, when the offered food pair does not contain the
food of number $j$,
$P(B_{j})=:q_j$ indicates the frequency of occurrence
of pair of food that does not contain food number $j$ .

By inspection of the Table \ref{Tabela1} of the functions $f_k$, $k=0,\ldots ,7$, we easily get the following relations:

\begin{eqnarray}\label{par}
P(C_{0}|B_{2})=P(\sum_{k=0}^{7} f_{k}(B_2)=0)=p_0+p_1+p_2+p_3\,,\nonumber\\
P(C_{0}|B_{1})=P(\sum_{k=0}^{7} f_{k}(B_1)=0)=p_0+p_1+p_4+p_5\,,\nonumber\\
P(C_{1}|B_{0})=P(\prod_{k=0}^{7} f_{k}(B_0)=1)=p_0+p_2+p_4+p_6\,,\\
P(C_{1}|B_{2})=P(\prod_{k=0}^{7} f_{k}(B_2)=1)=p_4+p_5+p_6+p_7\,,\nonumber\\\nonumber
P(C_{2}|B_{1})=P(\prod_{k=0}^{7} \frac{f_{k}(B_1)}{2}=1)=p_2+p_3+p_6+p_7\,,\\\nonumber\nonumber
P(C_{2}|B_{0})=P(\prod_{k=0}^{7} \frac{f_{k}(B_0)}{2}=1)=p_1+p_3+p_5+p_7\,,\\\nonumber
\end{eqnarray}
and $P(C_{0}|B_{0})=P(C_{1}|B_{1})=P(C_{2}|B_{2})=0$.\newline
The most valuable way of choosing the food by cat occurs
for such six conditional probabilities $(P(C_1|B_0),$ $P(C_2|B_0)$,$
  P(C_0|B_1)$,$P(C_2|B_1)$,
  $P(C_0|B_2)$,
  $P(C_1|B_2))$
  which fulfills the following condition:
  \begin{equation}
 \label{maximum}
 \omega_0=\omega_1=\omega_2=\tfrac{1}{3}.
 \end{equation}
Any six conditional probabilities, that for a fixed triple ($q_0,q_1,q_2$) fulfill (\ref{maximum})
will be called a \emph{cat's optimal strategy}. 
By simply calculation we obtain the relation between the optimal cat's strategy and frequencies
$q_j$ of appearance of food pairs (see \cite{r1}):
\begin{eqnarray}\label{solution}
q_2&=&\tfrac{1}{d}\bigg(\frac{P(C_0|B_1)+P(C_1|B_0)}{3}-P(C_0|B_1)P(C_1|B_0)\negthinspace\bigg),\nonumber\\
q_1&=&\tfrac{1}{d}\bigg(\frac{P(C_0|B_2)+P(C_2|B_0)}{3}-P(C_0|B_2)P(C_2|B_0)\negthinspace\bigg),\\
q_0&=&\tfrac{1}{d}\bigg(\frac{P(C_1|B_2)+P(C_2|B_1)}{3}-P(C_1|B_2)P(C_2|B_1)\negthinspace\bigg).\nonumber
 \end{eqnarray}
 The above relation defines a mapping $\mathcal{A}_0:D_3\rightarrow T_2$ of the three-dimensional cube ($D_3$) into a triangle ($T_2$) ( two-dimensional simplex,
 $q_0+q_1+q_2=1$ and $q_i\geq 0$), where $d$ is the determinant of the matrix of
parameters $P(C_j|B_k)$ (see \cite{r1}). The barycentric coordinates of a point of
this triangle are interpreted as probabilities $q_0, q_1$ i $q_2$. 

\section{Quantum modifications of Cat's Dilemma}
There are two obvious modifications of classical games \cite{r6}:
\begin{itemize}
	\item \textbf{prequantization} -- redefine the model by simply replacing the classical concepts of their equivalent quantum counterparts (reversal operation on qubits representing player's strategies).
	\item \textbf{quantization} -- reduce the number of qubit, allow arbitrary unitary transformation (at least one should not have its classical counterpart) preserving the fundamental property of classic game. Some modifications may lead to use the characteristic of quantum theory properties like measurement, entanglement, etc. 
\end{itemize}
We will treat the number of qubits as the main feature that distinguish prequantization from quantization.
\subsection{Prequantization}

Let us consider the three qubit system. All states of this system correspond to 8-dimensional Hilbert space ($H_8$).
The basis of $H_8$ will be denoted by 
 $\{|i\rangle\negthinspace\,\}$ $i=0,1,2,...,7$. 
 Let $|k\rangle\negthinspace$ denote the strategy of choosing the $f_k$ function.
 A family of convex vectors:
\begin{displaymath}
|\psi\rangle\negthinspace =\sum_{i=0}^{7}a_i|i\rangle\negthinspace,
\end{displaymath}
of norm 1 represents all cat's strategies spanned by the base vectors. Squares of coordinates moduli ($|a_i|^2$)
measure the probability of cat's making decision in choosing the $f_k$ function. 

Let $|a_i|^2=x_{2i}^2+x_{2i+1}^2$, $i=0,...,7$. Since $\sum_{i=0}^{7}|a_i|^2=1$, the model can be considered as 15-dimensional sphere:
\begin{displaymath}
\sum_{i=0}^{15}x_i^2=1.
\end{displaymath}
A similar calculation to that in classical model (we replaced $p_k$ by $|a_i|^2$) result in the projection $\mathcal{A}_1^p:S_{15}\rightarrow D_3$ of  15-dimensional sphere on to 3-dimensional cube of conditional probabilities:
\begin{align}\label{odwzpre}
P(C_{0}|B_{2})&=x_0^2+x_1^2+x_2^2+x_3^2+x_4^2+x_5^2+x_6^2+x_7^2,&\nonumber \\P(C_{1}|B_{0})&=x_0^2+x_1^2+x_4^2+x_5^2+x_8^2+x_9^2+x_{12}^2+x_{13}^2,\nonumber\\
P(C_{2}|B_{1})&=x_4^2+x_5^2+x_6^2+x_7^2+x_{12}^2+x_{13}^2+x_{14}^2+x_{15}^2, \\P(C_1|B_2)&=1-P(C_{0}|B_{2}),\nonumber\\
P(C_2|B_0)&=1-P(C_{1}|B_{0}),\nonumber &\\P(C_0|B_1)&=1-P(C_{2}|B_{1}).\nonumber
\end{align}
Combination of the above projection with (\ref{solution}) results in the projection $\mathcal{A}_p:S_{15}\rightarrow T_2$, $\mathcal{A}_p:=\mathcal{A}_0\circ\mathcal{A}_1^p$, 15 - dimensional sphere into a triangle $T_2$.

It is worth pointing out that the above quantum model has very similar properties to those of the classical model. It is the result of the same meaning of two parameters: $|a_i|^2$ in the quantum model and $p_k$ in the classical model. Such a simple replacement of the classical concept by a quantum equivalent is the main idea of prequantization. This is a good example of how the use of quantum models can simulate classic models. In the next part of the paper we explain the differences more accurately. Before this we will present a different way of modeling the game by limiting the number of qubits \cite{r2op}.


\subsection{Quantization}

Let us denote three different bases of two-dimensional Hilbert space as:
  $\{| 1 \rangle\negthinspace_{\,0}, | 2 \rangle\negthinspace_{\,0}\}$,
$\{| 0 \rangle\negthinspace_{\,1}$, $| 2 \rangle\negthinspace_{\,1}\}$,
$\{| 0 \rangle\negthinspace_{\,2}$, $| 1 \rangle\negthinspace_{\,2}\}=\{
(0,1)^{T},(1,0)^{T}\}$. The bases should be such that
bases 
\{$| 0 \rangle\negthinspace_{\,1}$, $| 2 \rangle\negthinspace_{\,1}$\},
\{$| 1 \rangle\negthinspace_{\,0}$, $| 2 \rangle\negthinspace_{\,0}$\} are the image of
\{$| 0 \rangle\negthinspace_{\,2}$, $| 1 \rangle\negthinspace_{\,2}$\} under
the transformations $H$
 and $K$ respectively : 
 \begin{displaymath}
\label{matrix maximal}
 H=\frac{1}{\sqrt{2}}\left(\begin{array}{cr}1 & 1 \\ 1 & -1 
\end{array}\right)\/,\quad
 K=\frac{1}{\sqrt{2}}\left(\begin{array}{cr}1 & 1 \\ i & -i 
\end{array}\right).
\end{displaymath}
These so-called conjugated
bases play  an important role in cryptography \cite{r27} and universality of quantum market games \cite{r28}.
Let us denote strategy of choosing the food number $k$, when the offered
food pair not contain the food of number $l$, as $| k \rangle\negthinspace_{\,l}$ ($k$, $l=0,1,2$, $k\ne l$).
 
A family $\{|z\rangle\negthinspace\}$ ($z \in \bar{\mathbb{C}}$) of convex vectors:
\begin{displaymath}
 | z \rangle\negthinspace:=| 0 \rangle\negthinspace_{\,2}+z
|1\rangle\negthinspace_{\,2}=| 0 \rangle\negthinspace_{\,1}+\frac{1-z}{1+z}
|2\rangle\negthinspace_{\,1}=| 1 \rangle\negthinspace_{\,0}+\frac{1+iz}{1-iz}
|2\rangle\negthinspace_{\,0}\,,
\end{displaymath}
defined by the parameters of the heterogeneous coordinates of
the projective space $\mathbb{C}P^{1}$, represents all quantum cat strategies
spanned by the base vectors.
The coordinates of the same strategy
$| z \rangle\negthinspace$\, read (measured) in various bases define quantum cat’s
preferences toward a food pair represented by the base vectors. Squares of their moduli, after normalization, measure the conditional
probability of quantum cat's making decision in choosing a particular product, when the choice is related to the suggested
food pair.
Strategies $|z\rangle\negthinspace$ can be parameterized by the sphere $(x_1,x_2,x_3)\in S_2\backsimeq \overline{\mathbb{C}}$. We get the mapping
$\mathcal{A}_1^q:S_2\rightarrow D_3$ of 2-dimensional sphere $S_2$ onto the 3-dimensional cube of conditional
probabilities:
\begin{eqnarray}
P(C_0|B_2)=\frac{1-x_3}{2}, \qquad P(C_1|B_2)=\frac{1+x_3}{2},\nonumber \\
P(C_0|B_1)=\frac{1+x_1}{2},\qquad P(C_2|B_1)=\frac{1-x_1}{2}, \\
P(C_1|B_0)=\frac{1+x_2}{2},\qquad P(C_2|B_0)=\frac{1-x_2}{2}.\nonumber  
	\end{eqnarray}
 Combination of the above projection with (\ref{solution}) results in the projection $\mathcal{A}_q:S_2\rightarrow T_2$, $\mathcal{A}_q:=\mathcal{A}_0\circ\mathcal{A}_1^q$  
of 2-dimensional sphere $S_2$ into a triangle $T_2$. 

 \section{The evolution of the model}
 In this Section we compare the effects offered by various models of the problem. We present a range of representations $\mathcal{A}_0$, $\mathcal{A}_p$, $\mathcal{A}_q$ for 10 000 randomly selected points with respect to constant probability distribution in the space of strategy. In the classical case, justification of such equipartition of probability may be found in Laplace's principle of insufficient reason \cite{r31}. In quantum models constant probability distribution corresponds to the Fubini-Study measure \cite{r37}.
 
 \subsection{Optimal strategies}
\begin{figure}[htbp]
         \centering{
       \includegraphics[
          height=1.25in,
          width=1.4in]%
         {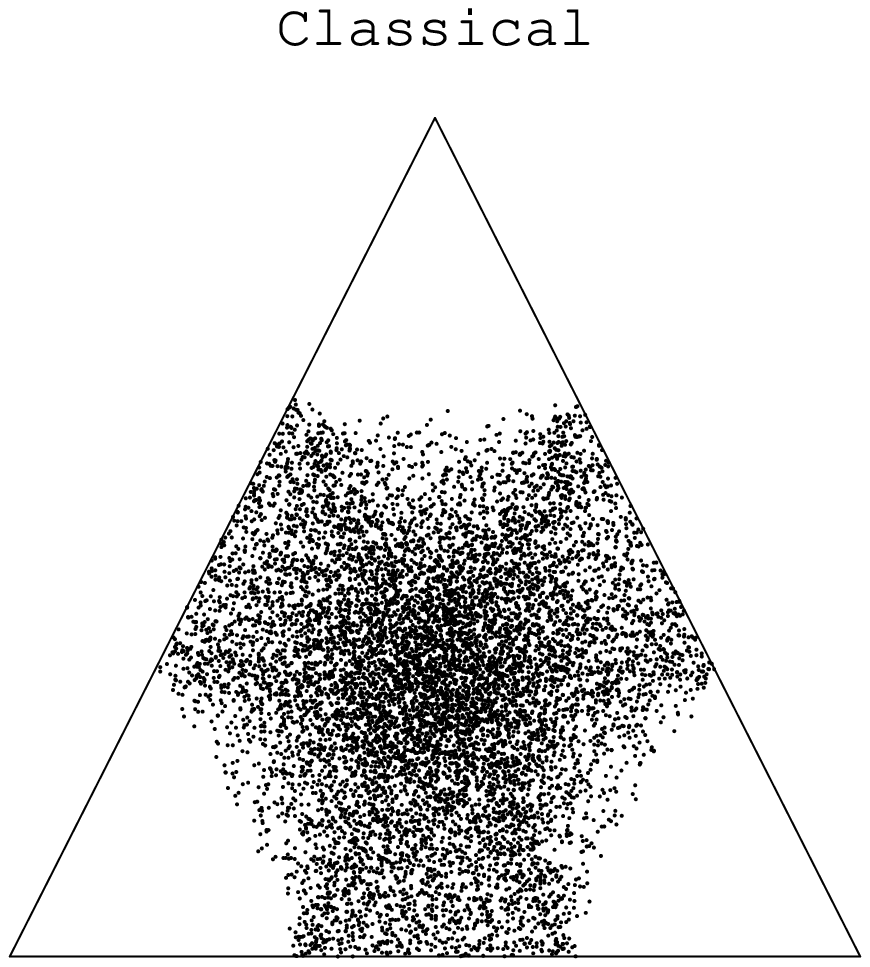}
       \includegraphics[
          height=1.25in,
          width=1.4in]%
         {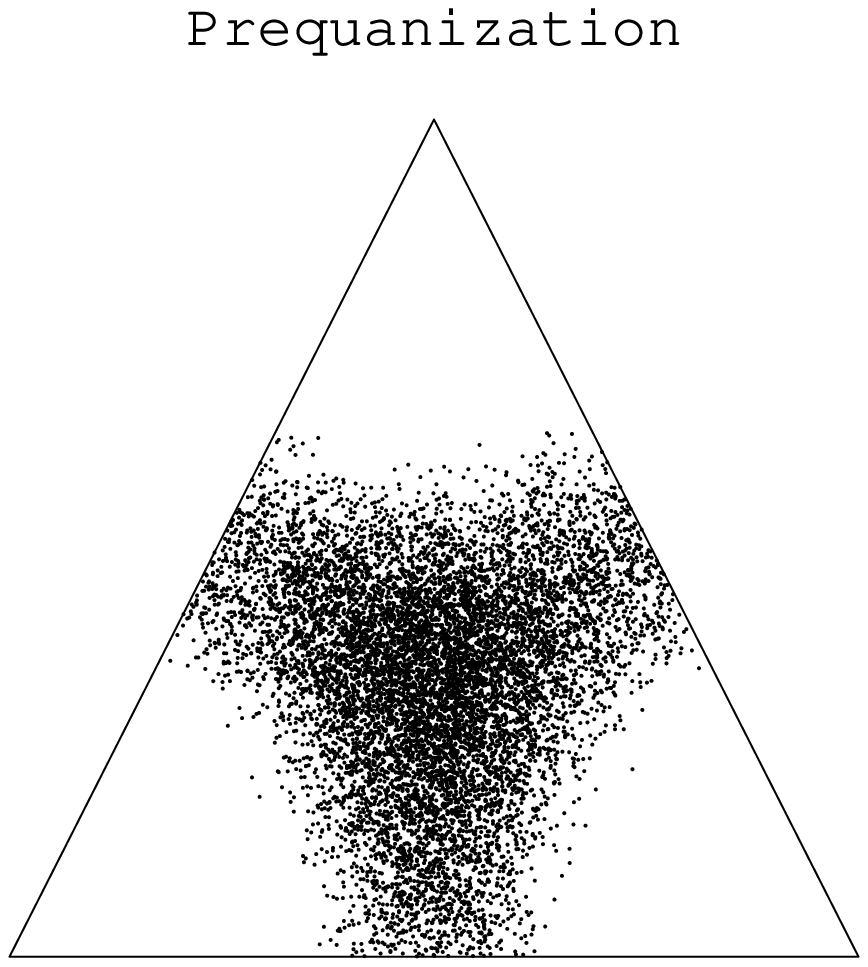}
         \includegraphics[
                      height=1.25in,
                    width=1.4
                      in]%
       {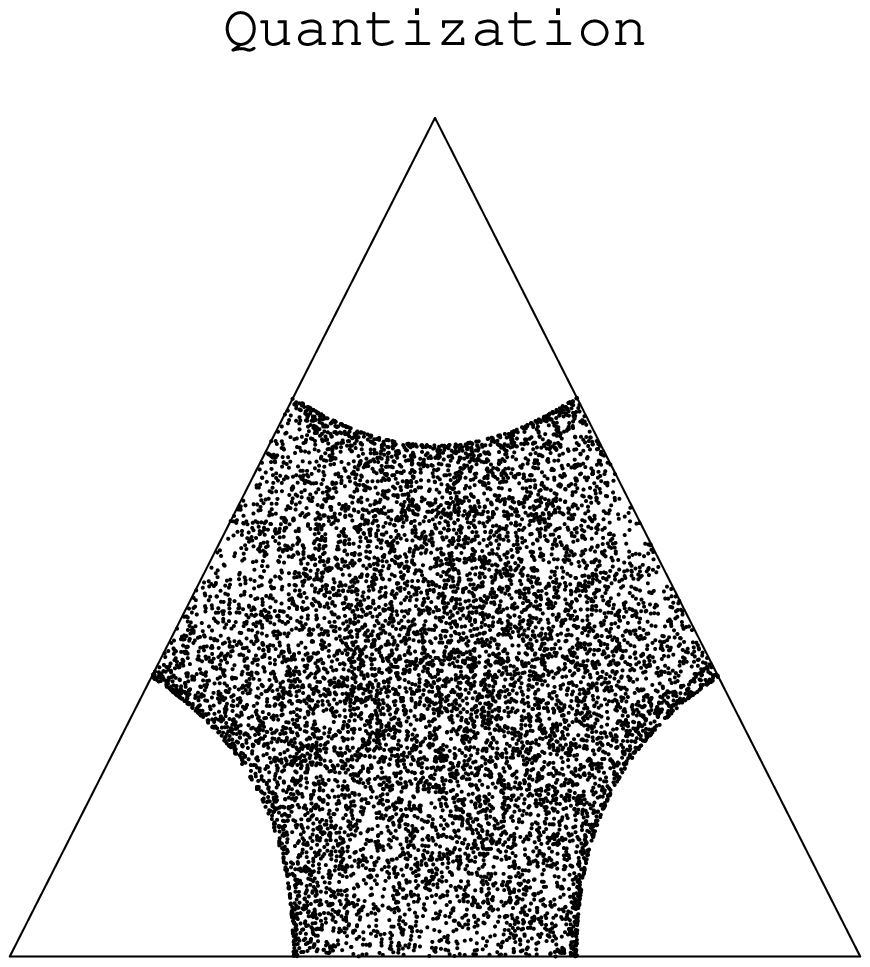}
       }\caption{Optimal strategies.}
         \label{qhex}
\end{figure}

 Fig. \ref{qhex} on page \pageref{qhex} presents the areas (in each model) of frequency $q_m$ of appearance of individual choice alternatives between two types of food, for which optimal strategies exist.
The biggest part of the triangle is covered in the classical model and quantum 8-dimensional model. In these cases, these areas overlap, because the mapping (\ref{odwzpre}) is surjective (so the areas in a 8-dimensional model are reduced to the classical case).
In the 2-dimensional  quantum case the area of the simplex corresponding to the optimal strategies has become slightly diminished in relation to the classical model and 8-dimensional quantum model. It is also worth mentioning that in the classical model and 8-dimensional quantum model we deal with a sort of condensation of optimal strategies in the central part of the picture in the area of the balanced frequencies of all pairs of food. In the 2-dimensional  quantum case they are more evenly spread out, although they also appear less frequently towards the sides of the triangle.


 \subsection{Transitive and intransitive strategies}
\begin{figure}[htbp]
          \centering{
        \includegraphics[
           height=1.25in,
           width=1.4in]%
          {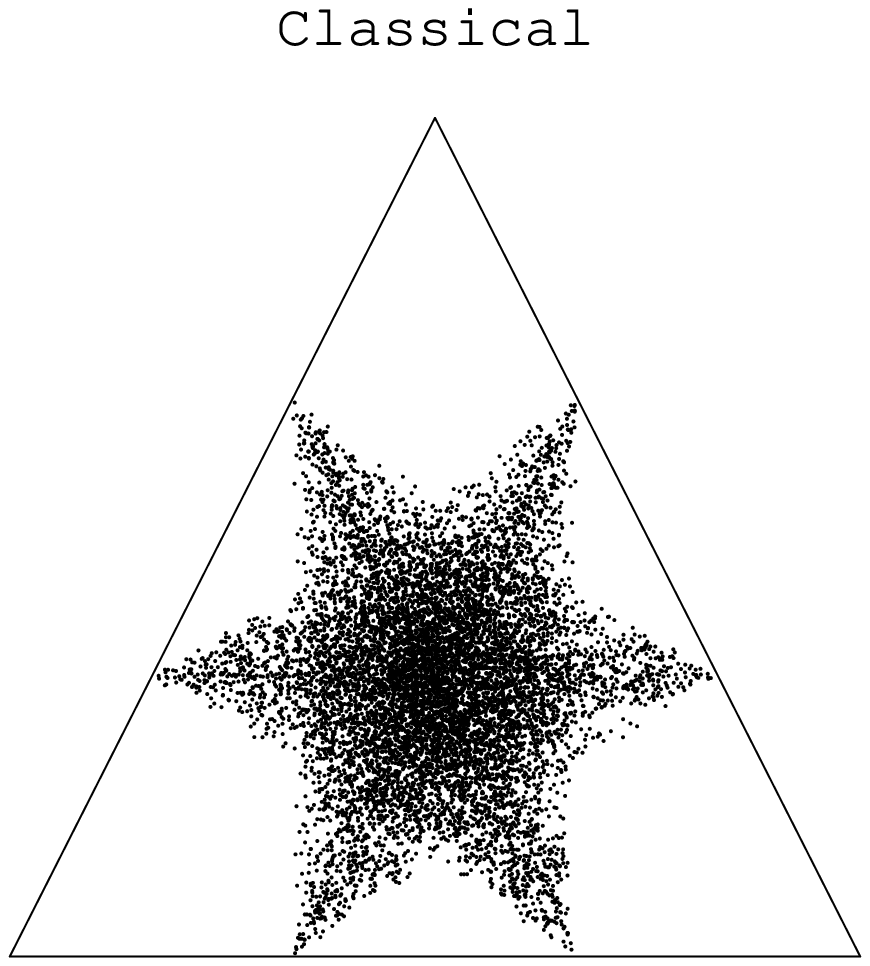}
        \includegraphics[
           height=1.25in,
           width=1.4in]%
          {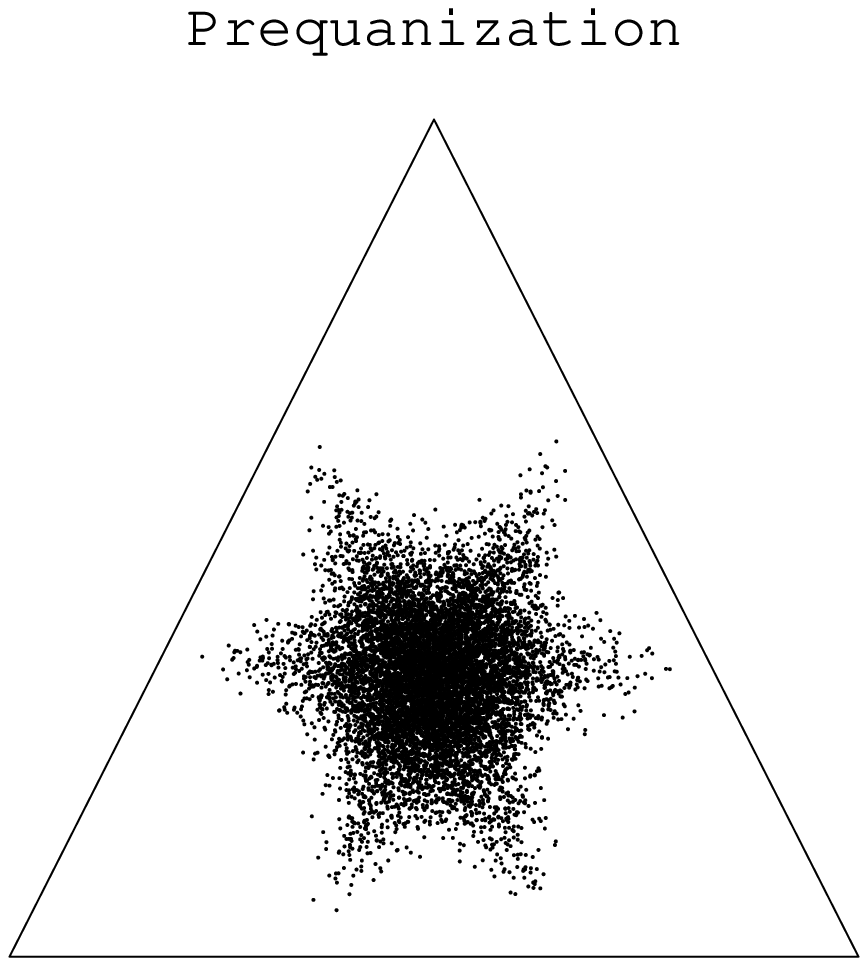}
          \includegraphics[
                       height=1.25in,
                       width=1.4
                       in]%
                       {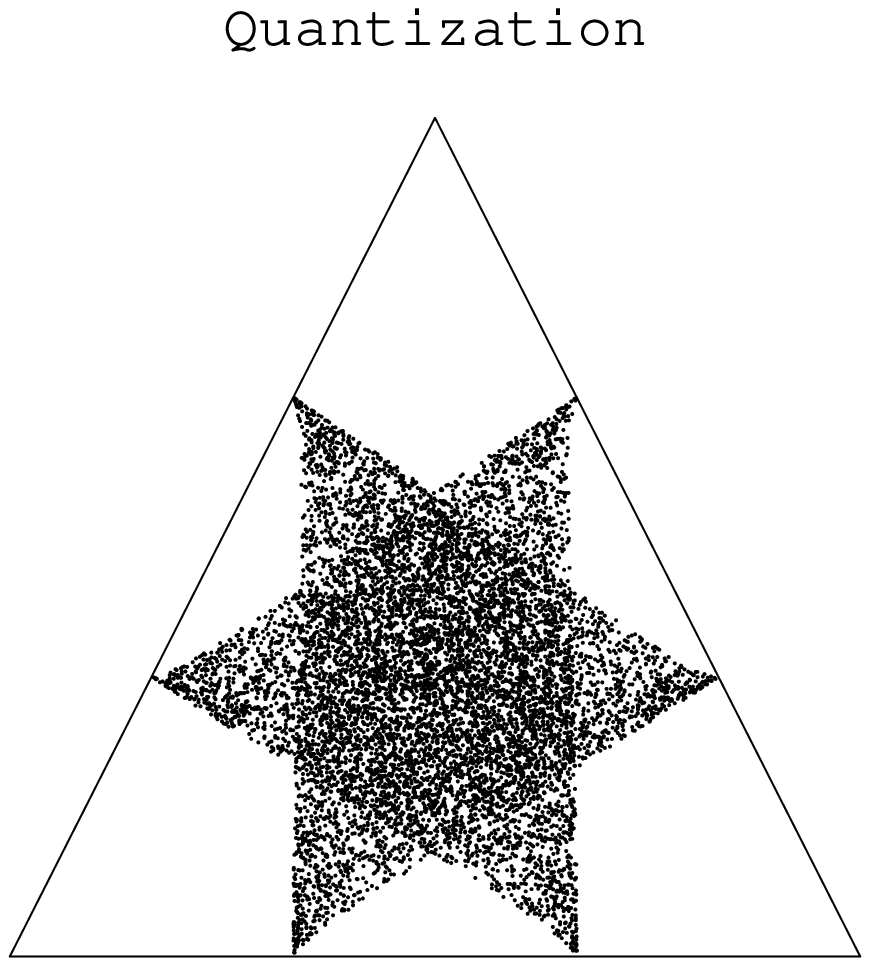}
          }\caption{Optimal intransitive strategies.}
          \label{qstar}
\end{figure}

We deal with an intransitive choice if one of the following conditions is fulfilled:
 \begin{itemize}
  \item $P(C_2|B_1)<\frac{1}{2}$, 
  $P(C_1|B_0)<\frac{1}{2}$, 
  $P(C_0|B_2)<\frac{1}{2}$\,,
  \item $P(C_2|B_1)>\frac{1}{2}$, 
  $P(C_1|B_0)>\frac{1}{2}$, 
  $P(C_0|B_2)>\frac{1}{2}$\,.
 \end{itemize}
 It may be seen in Fig. \ref{qstar} on page \pageref{qstar} in what part of the simplex of parameters $(q_0$, $q_1$, $q_2)$  intransitive strategies may be used in each model. They form a six-armed star composed of two triangles (any of them corresponding to one of two possible intransitive orders). As in the case of optimal strategies, one can notice that the 2-dimensional quantum variant is characterized by higher regularity, the star has clearly marked boundaries. A common feature of all models is that the intransitive optimal strategies often occur in the center of the simplex (near point $q_0 = q_1 = q_2 = \frac{1}{3}$).

\begin{figure}[htbp]
          \centering{
        \includegraphics[
           height=1.25in,
           width=1.4in]%
          {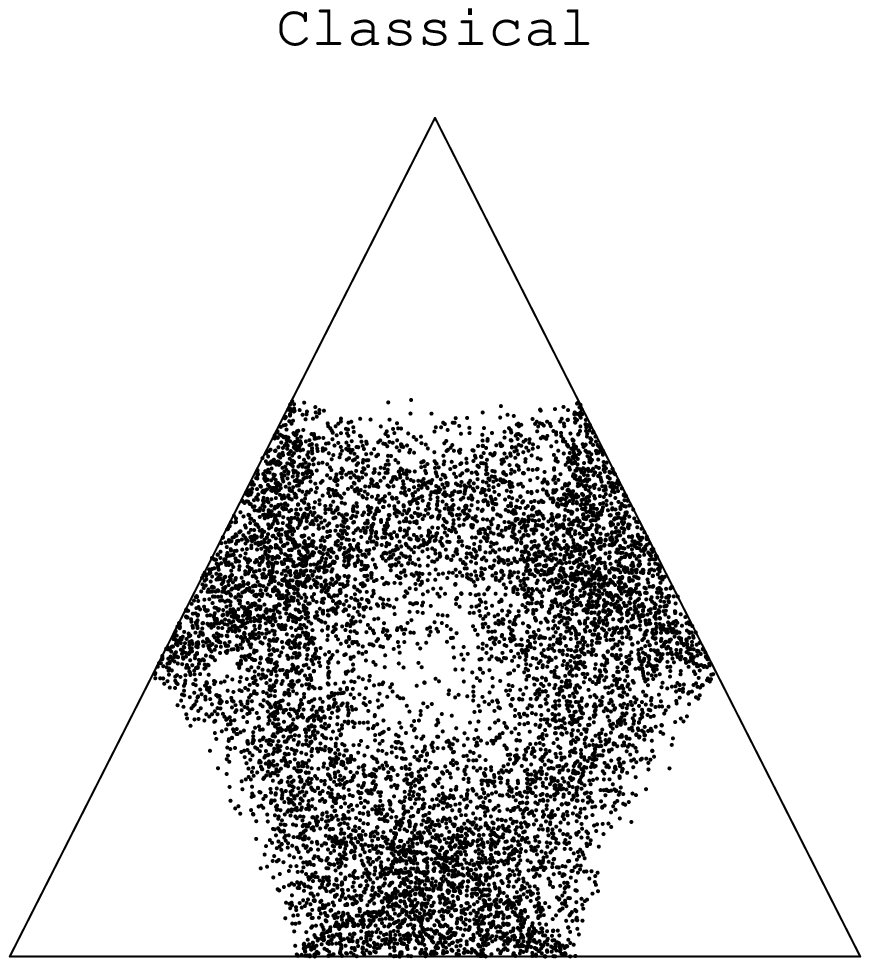}
        \includegraphics[
           height=1.25in,
           width=1.4in]%
          {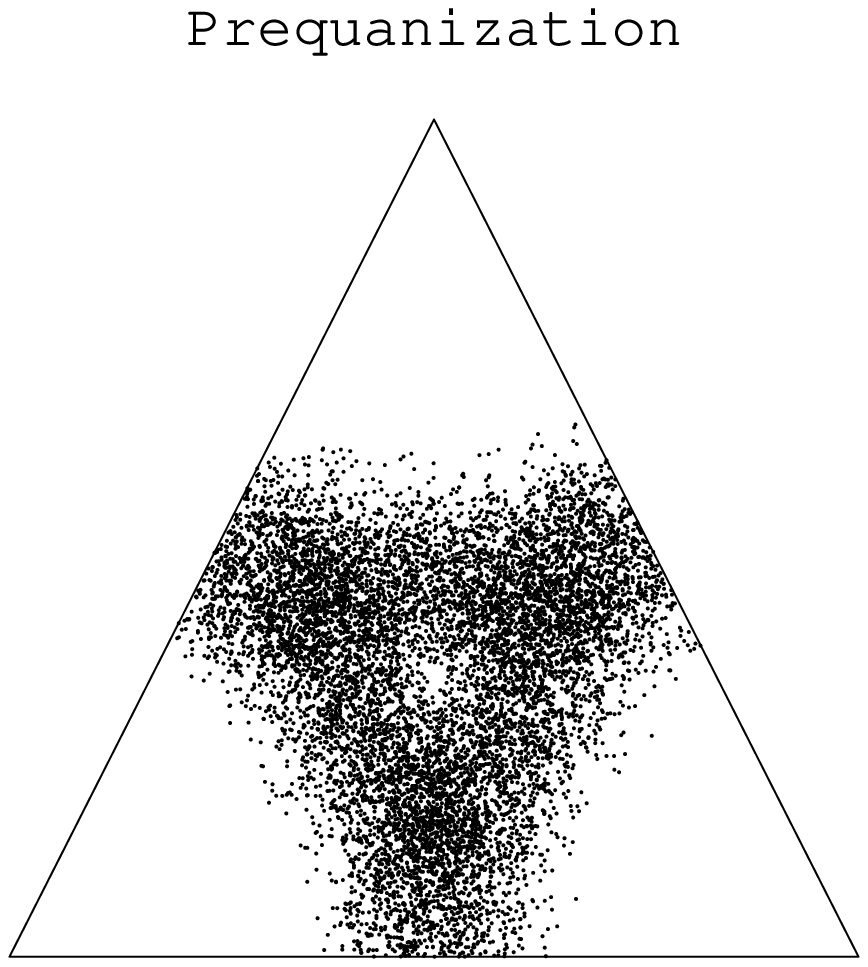}
          \includegraphics[
                       height=1.25in,
                       width=1.4in]%
        {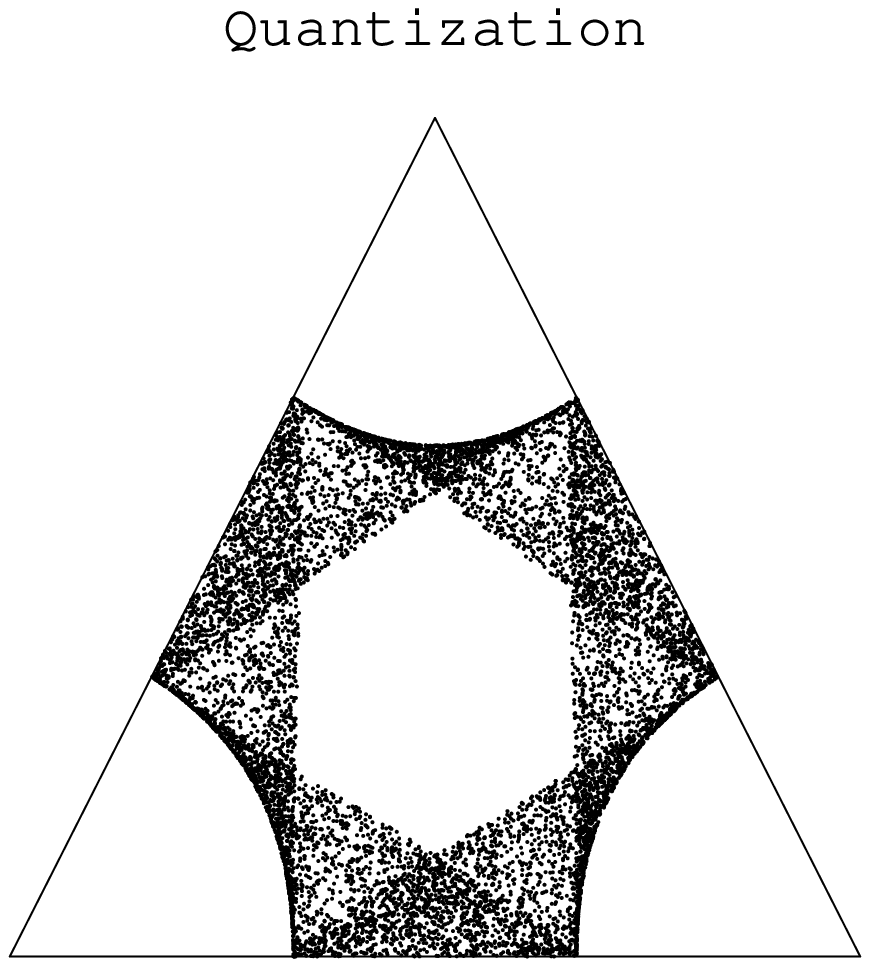}
          }\caption{Optimal transitive strategies. }
        \label{qtrans}
\end{figure}

Fig. \ref{qtrans} on page \pageref{qtrans} presents a simplex area for which there exist transitive optimal strategies in each model. In both the classical and 8-dimensional quantum model, transitive strategies cover the same area of the simplex as all optimal strategies (but they occur less often in the center of the simplex where intransitive strategies occur more often). The 2-dimensional quantum model is essentially different -- transitive optimal strategies do not appear within the boundaries of the hexagon in the central part of the triangle. Therefore, a transitive order whose working effects are identical cannot be defined for each intransitive order. The 2-dimensional quantum model gives a considerable weight to intransitive orders -- for some frequencies of appearance of pairs of food, the quantum cat is able to achieve optimal results only thanks to the intransitive strategy.  

To make the analysis clear, we sum up our quantitative results in Table  \ref{Tabela2} on page \pageref{Tabela2} \cite{r1, r2op}:
\section{Electoral interpretation}
To the model analyzed above there can be given an electoral interpretation, referring to the reflections of Condorcet.


Let's suppose that the elections concern three candidates(no 0, no 1, no 2), the decisions of the voters are one - bite  (I support - I don't support) and their preferences are formed optimally, i.e. they are divided into 3 parts of an identical measure ($\omega_0=\omega_1=\omega_2=\tfrac{1}{3}$)\footnote{It is analogous to the most efficient tactics in the 20 questions game.}. Let's mention at least the presidential election in the USA between G.W Bush and A. Gore, which brought Bush a minimal victory. Also the parliamentary elections in Italy and Germany were characterized by very equalized results of the competed parties. From the theoretical point of view, this situation is extremely interesting, because the smallest fluctuations of the electoral preferences influence considerably the elections result. It is worth looking at the model analysis presented in the paper also in this context.  \\
The proposed electoral interpretation is characterized by a limited possibility of choice.  At first, let's draw a pair with the probabilities $q_0, q_1, q_2$ \footnote{This phase can be associated with the first ballot, when two candidates with the biggest support come to the next phase -- the second ballot. It is a characteristic of elections in many countries, mainly in the European ones.} and then we choose a candidate from the proposed pair. We see that when the probabilities of appearance of the respective pairs are also balanced (they oscillate more or less in the middle part of the triangle), then the decision of the voters can be intransitive. In the 2-dimensional quantum model decision is exclusively intransitive. Are the preferences of the voters intransitive, when in the elections the candidates have equal chances for victory? Which model gives the reality better? It is worth verifying similar hypotheses, because it will allow to evaluate in a better way the utility of the quantum models. \\
In the quantum case ``the collective voter'' -- the electorate has a completely clear, and what's more, a pure electoral strategy. This strategy could come into being as a superposition of the strategies of single voters interfering with each other. 
It is a paradoxical extension of possibilities of intransitive preferences to a single voter. 

There are not many publications dealing with the electoral issue in the area of quantum theory of games. This is a new and a very attractive scope of investigations. It is possible that the quantum electoral allegories deliver the specific character of the elections in a better way. 


\section{Quatization -- Prequatization. Is this useful?}
In the paper we discussed two different approaches to modeling a simple sequential game in the quantum game theory language. We see that the 8-dimensional model (prequantization) gave similar results to the classical model. The important difference is in the assumptions. In the classical model we based our assumptions on Laplace's principle of insufficient reason \cite{r31}. Quantum models do not have this defect. The measurement method for quantum cat's strategy is justified by the fact that constant probability distribution corresponds to the Fubini-Study measure, which is the only invariant measure in relation to any change of quantum cat's decision regarding the chosen strategy. As a disadvantage of prequantization can be the fact that the frequencies $q_m$, for which there are given optimal algorithms of a different type, concentrate too much on point $q_0=q_1=q_2=\tfrac{1}{3}$. This is probably a consequence of the big dimension of the sphere from which the points are drawn. 

It is worth pointing out that quantum models give very different effects. Intransitive orders have greater importance in the 2-dimensional model (quantization) than in the 8-dimensional model (and classical one) and we receive this distinction in one language (the quantum game theory).
Perhaps this will allow to model more complex situations requiring coupling between the different characteristics (in our case between intransitivity and transitivity).


It may be recalled that such distinctions have been known in the social sciences for a long time. The divide between the consciousness and unconsciousness, or Freud's personality theory (Freud believed that personality has three structures: the Id, the Ego, and the Superego) are important points of theory which try to describe human behavior.

In work \cite{r26} the authors, recognizing the similarities between the structure of computer operating systems (distinguish the shell and the kernel) and quantum game that is perceived as an algorithm implemented by a specific
quantum process, they have proposed Quantum Game Model of Mind. They compared the division into kernel and shell within the quantum game with the psyche division into consciousness and unconsciousness. 
Kernel means Ego, so the conscious part of the psyche, which is often described as rationality. It controls the cognitive and intellectual functions. 
Shell means Id, so the unconscious part of the personality. It acts as an impulse principle, an immediate fulfilling of needs (pleasure principle). Id is a human biological sphere. It is quite similar in case of transitivity and intransitivity. 

As mentioned before, the transitive preferences are generally considered as one of the components of a rational behavior.  During the evolution of the model  \emph{Cat's dilemma} (from the classical one through the prequantization) the transitive preferences dominate the intransitive ones. In case of our example the prequantization shows better the behaviors, which can be associated with Ego, so generally considered as rational.
On the other hand, the quantization gives the predominance to the intransitive preferences \footnote{For some frequencies of appearance of
pairs of food, quantum cat is able to achieve optimal results only thanks to the intransitive strategy.}, often shown spontaneously, under the influence of an impulse. It can be associated with Id.

In the work there was only discussed a simple model of behaviors and it is difficult to prejudge, that the above mentioned conformity, noticeable in our example, in the context of intransitive and transitive preferences, can be seen in case of other or more complicated models of behaviors. Therefore, it is characteristic, that the phenomena occurring in the nature are both transitive and intransitive. Our decisions are also based on both transitive and intransitive preferences, depending on the concrete situation. It is possible that this duality can be investigated throughout different constructions of quantum models, parallel to the way presented in the paper, whereas their variety will allow to describe in a better way the surrounding reality and it can make up important completion of classic theories.   

Quantum mechanics caused a huge breakthrough in the contemporary physics and it can make such a breakthrough also in other disciplines. 
Who knows, if we won't find in the language of the quantum games theory a good method for simulating a human thought process, which doesn't have to show a biological specific character of brain (similarly, the airplanes don't flutter their wings). It is worth considering the quantum games in this context, keeping in mind, that even in their classic form they were used for describing intelligent behaviors.

It is too simple to recognize the transitivity as an inseparable component of the rational behavior. In the example presented in the research the cat, thanks to the intransitive preference, secured the completeness of food needed for maintaining health. In the quantization model it has such a possibility (for some frequencies $q_m$)  only thanks to the intransitive strategies. At this stage it is difficult to evaluate, if this increase of intransitivity in the quantum models has a more general character. It is possible that they give better our spontaneously shown preferences, which often have this feature. 

Two different ways of quantum modeling of game gave significantly different results. The model of prequantization gives more importance to the transitive strategies, whereas the model of quantization to the intransitive ones. Will such distinction in the way of the quantum description of game allow to catch various features characterizing a given problem in more general and complex situations? This question still remains open.

\newpage
\section*{Figures}
\newpage
\section*{Tables}
\begin{table}[htbp]
\caption{The table defining all possible choice functions
$f_k$.}\vspace{1ex} \centering\footnotesize
\begin{tabular}{c c c c c c c c c} \hline
\vphantom{$\int^1$} function \vphantom{$F^K$}$f_k$: &
${f_{0}}$ & ${f_{1}}$ & ${f_{2}}$ & ${f_{3}}$ & ${f_{4}}$ &
${f_{5}}$ & ${f_{6}}$ & ${f_{7}}$\\[1pt]
\hline
\vphantom{$\int^1$}$f_{k}(1,0)=$\vphantom{$F^K$} & 0 & 0 & 0 & 0 & 1 & 1 & 1 & 1
\\[1pt] \vphantom{$\int^1$}$f_{k}(2,0)=$\vphantom{$F^K$}
 & 0 & 0 & 2 & 2 & 0 & 0 & 2 & 2 \\[1pt]\vphantom{$\int^1$}
$f_{k}(2,1)=$ \vphantom{$F^K$}& 1 & 2 & 1 & 2 & 1 & 2 & 1 & 2
\\[1pt]\hline \vphantom{$\int^1$}frequency $p_k$: & $p_0$ & $p_1$ & $p_2$ & $p_3$ & $p_4$
& $p_5$ & $p_6$ & $p_7$
\\[1pt]\hline
\end{tabular}\label{Tabela1}
\end{table}
\begin{table}[htbp]
\caption{Comparison of achievability of various types of optimal strategies in models.}
\vspace{1ex} \centering\footnotesize
\begin{tabular}{c c c c} \hline
\vphantom{$\int^1$}&
 All &  Intransitive &  Transitive
\\[1pt]\hline
\vphantom{$\int^1$}Classical and prequantization \vphantom{$F^K$} & 67 $\%$ & 44 $\%$ & 67 $\%$ 
\\[1pt] \vphantom{$\int^1$}Quantization\vphantom{$F^K$}
 & 60 $\%$  & 44 $\%$ & 37 $\%$ \\[1pt]\hline
\end{tabular}\label{Tabela2}
\end{table}

\begin{thebibliography}{00}
\bibitem{r1}  M. Makowski, E. W. Piotrowski,  Fluctuat. Noise Lett. 5 (2005) L85.

\bibitem{r2} J. Eisert, M. Wilkens, and M. Lewenstein, Phys. Rev. Lett.  83 (1999) 3077.

\bibitem{r3} D. Meyer, Phys. Rev. Lett. 8 (1999) 1052.

\bibitem{r4} A. P. Flitney, D. Abbott,  Fluctuat. Noise Lett.  2  (2002) R175.

\bibitem{r5} E. W. Piotrowski, J. S\l adkowski,  Int. J. Theor. Phys.  42 (2003) 1089.

\bibitem{r6} E. W. Piotrowski, J. S\l adkowski, in: C. V. Benton (Ed.), Mathematical Physics Research at the Cutting Edge, Ed. , Nova Science, New York, 2004 pp.247-268.


\bibitem{r7} A. C. Elitzur , L. Vaidman, Found. Phys.  23 (1993) 987. 

\bibitem{r8} L. Vaidman, Found. Phys.  29 (1999) 615.

\bibitem{r9} A. Iqbal, A. H. Toor,  Phys.Lett.A   280 (2001) 249.

\bibitem{r10}  A. Iqbal, A. H. Toor, Phys.Lett.A 294 (2002) 261.

\bibitem{r3Niels} M. A. Nielsen, I. L. Chuang,  Quantum Computation and Quantum information, Eds.Cambridge University Press, 2000.

\bibitem{r29} M. Buchanan, Variety really is the spice of life, New Scientist, 21 Aug. 2004.

\bibitem{r33}  L. Grosenick, T. S. Clement, R. D. Fernald,  Nature  445 (2007) 429.

\bibitem{r23} R. M. Dawes,  in: D. Gilbert, S. Fiske and G. Lindzey (Eds), The Handbook of Social Psychology, Vol. 1  Boston, MA: McGraw-Hill, 1998, pp. 497-548.

\bibitem{r18} C. Hauert, S. De Monte, J. Hofbauer,  K. Sigmund,  J. theor. Biol.  218 (2002) 187.

\bibitem{r19} D. Semmann, H.J. Krambeck  M. Milinski,  Nature  425 (2003) 390.

\bibitem{r24} K. J. Arrow,  Social Choice and Individual Values, Yale Univ. Press, New York, 1951.

\bibitem{r34} M. Gardner,   Mathematical Games: The Paradox of the Nontransitive Dice and the Elusive Principle of Indifference, Sci. Amer. 223,  Dec. 1970. 

\bibitem{r35} W. Penney, {\em A Waiting Time Problem}, Journal of Recreational Mathematics, October 1969.

\bibitem{r15} L. Boddy, FEMS Microbiology Ecology  31, (2000) 185.

\bibitem{r16} S. Shafir, Animal Behaviour  48 (1994) 55. 

\bibitem{r30} A. Poddiakov, J. Valsiner, in: In L. Rudolph,  J. Valsiner (Eds.),  Qualitative mathematics for the social sciences, London: Psychology press (in press).

\bibitem{r22} P. C. Fishburn, J. Risk and Uncertainty 4 (1991) 113.

\bibitem{r2op} M. Makowski, E. W. Piotrowski,  Phys.Lett.A 355 (2006) 250.

\bibitem{r36} J. B. Busemeyer, Z. Wang, J. T. Townsend, J. Math. Psychology  50 (2006) 220.

\bibitem{r25} H. Steinhaus,  Memoirs and Notes, Aneks, London, (1992) in Polish.

\bibitem{r27} S. Wiesner,  Conjugate coding,  Sigact News  15 (1983) 78.

\bibitem{r28} I. Paku\l a, E.W. Piotrowski, J. S\l adkowski, Physica A  385 (2007) 397.

\bibitem{r31} P. Dupont, Rend. Sem. Mat. Univ. Politec. Torino, 36 (1977/78) 125.

\bibitem{r37} M. Berger,  Geometry, Springer-Verlag, Berlin, 1987.


 

\bibitem{r26} K. Miakisz, E. W. Piotrowski, J. S\l adkowski, 
Theoret. Comput. Sci. 358  (2006) 15. 


\end{thebibliography}
\end{document}